\documentclass{article}
\usepackage{spconf}
\newcommand{\figsize}{3.2}
\usepackage{graphicx,psfrag}
\usepackage{url}
\usepackage[usenames]{color}
\usepackage[cmex10]{amsmath}
\usepackage{amsfonts,amssymb,latexsym,cite,color,multirow,rotating}
\usepackage{algorithm,algorithmic}
\usepackage{bbm}
\usepackage{hyperref}
\usepackage{enumitem} % easy list customization

 \newcommand{\putTable}[3]{\begin{table}[t!]
                            \centering
                            #3
                            \vspace{-3 mm}
                            \caption{#2}
                            \label{tab:#1}
                            \vspace{-3 mm}
                          \end{table} }
 \newcommand{\putFrag}[4]{\begin{figure}[t!]
                            \centering
                            #4
                            \includegraphics[width=#3in]{figures/#1.eps}
                            \vspace{-2mm}
                            \caption{#2}
                            \label{fig:#1}
                          \vspace{-3 mm}
                          \end{figure}}

 \renewcommand{\tilde}{\widetilde}
 \renewcommand{\hat}{\widehat}
 \newcommand{\defn}{\triangleq}

 \newcommand{\tvec}[1]{\ensuremath{\Tilde{\boldsymbol{#1}}}}
 \newcommand{\ovec}[1]{\ensuremath{\overline{\boldsymbol{#1}}}}
 \newcommand{\hvec}[1]{\ensuremath{\Hat{\boldsymbol{#1}}}}
    
 \renewcommand{\vec}[1]{\ensuremath{\boldsymbol{#1}}}
 \newcommand{\mat}[1]{\ensuremath{\begin{bmatrix}#1\end{bmatrix}}}

 \newcommand{\norm}[1]{\ensuremath{\| #1 \|}}
 \newcommand{\mc}[1]{\ensuremath{\mathcal{#1}}}
 \newcommand{\st}{{~\text{s.t.}~}}

 \newcommand{\Real}{{\mathbb{R}}}

 \newcommand{\herm}{^\textsf{H}}

 \DeclareMathOperator{\sgn}{sgn}
 
 \DeclareMathOperator{\E}{E}
 \DeclareMathOperator{\var}{var}

 \DeclareMathOperator*{\argmax}{arg\, max}
 \DeclareMathOperator*{\argmin}{arg\, min}

 \renewcommand{\eqref}[1]{(\ref{eq:#1})}

 \newcommand{\figref}[1]{Fig.~\ref{fig:#1}}
 \newcommand{\figsref}[1]{Figs.~\ref{fig:#1}}
 \newcommand{\tabref}[1]{Table~\ref{tab:#1}}
 
 \newcommand{\secref}[1]{Section~\ref{sec:#1}}

 \newcounter{comment}[section]
 
 \newcounter{texthead}[section]

 \newcommand{\SNR}{\textsf{SNR}}
 \newcommand{\MSE}{{\mathsf{MSE}}}
 \newcommand{\MMSE}{\mathsf{MMSE}}
 \newcommand{\NMSE}{\textsf{NMSE}}
 \newcommand{\MAP}{\mathsf{MAP}}
 \newcommand{\bethe}{\textrm{Bethe}}

 \newcommand{\X}{\textsf{x}}
 
 \newcommand{\A}{\textsf{a}}
  
  \newcommand{\mA}{\textbf{\textsf{A}}}
 \newcommand{\Y}{\textsf{y}}
 \newcommand{\Z}{\textsf{z}}

 \newcommand{\U}{\textsf{u}}
  \newcommand{\mU}{\textbf{\textsf{U}}}
 \newcommand{\V}{\textsf{v}}
  \newcommand{\mV}{\textbf{\textsf{V}}}
 \newcommand{\vX}{\textsf{\textbf{\textit{x}}}}
  
 \newcommand{\vY}{\textsf{\textbf{\textit{y}}}}
 
 \newcommand{\vZ}{\textsf{\textbf{\textit{z}}}}

\title{Adaptive Damping and Mean Removal for\\the Generalized Approximate Message Passing Algorithm}
\name{
Jeremy Vila$^{\star}$ 
\quad Philip Schniter$^{\star}$ 
\quad Sundeep Rangan$^{\dagger}$ 
\quad Florent Krzakala$^{\ddagger}$
\quad Lenka Zdeborov{\'a}$^{\circ}$
\thanks{This work has been supported in part by NSF grants IIP-0968910, CCF-1018368, and CCF-1218754, an allocation of computing time from the Ohio Supercomputer Center, and by European Union’s 7th Framework Programme (FP/2007-2013)/ERC Grant Agreement 307087-SPARCS.}
}
\address{
$^{\star}$ Dept. of ECE, The Ohio State University, Columbus, OH 43202, USA.\\
$^{\dagger}$ Dept. of ECE, Polytechnic Institute of New York University, Brooklyn, NY 11201, USA.\\
$^{\ddagger}$ Sorbonne Universit{\'e}s, UPMC Univ Paris 06 and {\'E}cole Normale Sup{\'e}rieure, 75005 Paris, France.\\
$^{\circ}$ Institut de Physique Th{\'e}orique, CEA Saclay, and CNRS URA 2306, 91191 Gif-sur-Yvette, France.
}

\begin{document}
\ninept
\maketitle
\setlength{\arraycolsep}{0.5mm}

\begin{abstract}
The generalized approximate message passing (GAMP) algorithm is an efficient method of MAP or approximate-MMSE estimation of $\vec{x}$ observed from a noisy version of the transform coefficients $\vec{z}=\vec{Ax}$.
In fact, for large zero-mean i.i.d sub-Gaussian $\vec{A}$, GAMP is characterized by a state evolution whose fixed points, when unique, are optimal.
For generic $\vec{A}$, however, GAMP may diverge.
In this paper, we propose adaptive-damping and mean-removal strategies that aim to prevent divergence.
Numerical results demonstrate significantly enhanced robustness to non-zero-mean, rank-deficient, column-correlated, and ill-conditioned $\vec{A}$.%
\end{abstract}


%%%%%%%%%%%%%%%%%%%%%%%%%%%%%%%%%%%%%%%%%%%%%%%%%%%%%%%%%%%%%%%%%%%%%%%%%%%%%
\section{Introduction} \label{sec:intro}

Consider estimating a realization $\vec{x}\in\Real^N$ of a random vector $\vX$ with statistically independent components $\X_n\sim p_{\X_n}$ from observations $\vec{y}=[y_m]\in\Real^M$ that are conditionally independent given the transform outputs 
\begin{equation}
\vZ = \vec{A}\vX 
\label{eq:z}
\end{equation}
for some known matrix $\vec{A}=[a_{mn}]\in\Real^{M\times N}$.
Here, the likelihood function can be written as $p_{\vY|\vZ}(\vec{y}|\vec{Ax})$ with separable $p_{\vY|\vZ}$, i.e., $p_{\vY|\vZ}(\vec{y}|\vec{z})=\prod_{m=1}^M p_{\Y_m|\Z_m}(y_m|z_m)$.
Such problems arise in a range of applications including statistical regression, inverse problems, and compressive sensing.
Note that, for clarity, we use san-serif fonts (e.g., $\vX,\X_n$) to denote random quantities and serif fonts (e.g., $\vec{x},x_n$) to denote deterministic ones.

Assuming knowledge of the prior $p_{\vX}(\vec{x})=\prod_{n=1}^N p_{\X_n}(x_n)$ and likelihood $p_{\vY|\vZ}(\vec{y}|\vec{z})$, typical estimation goals are to compute the minimum mean-squared error (MMSE) estimate $\hvec{x}_\MMSE\defn\int_{\Real^N} \vec{x} p_{\vX|\vY}(\vec{x}|\vec{y})d\vec{x}$ or the maximum a posteriori (MAP) estimate $\hvec{x}_\MAP\defn\argmax_{\vec{x}} p_{\vX|\vY}(\vec{x}|\vec{y})=\argmin_{\vec{x}} J_\MAP(\vec{x})$ for the MAP cost
\begin{equation}
J_\MAP(\hvec{x})\defn -\ln p_{\vY|\vZ}(\vec{y}|\vec{A}\hvec{x}) - \ln p_{\vX}(\hvec{x}).
\label{eq:Jmap}
\end{equation}
Recently, the \emph{generalized approximate message passing} (GAMP) algorithm \cite{Rangan:ISIT:11} has been proposed as a means of tackling these two problems in the case that $M$ and $N$ are large. 
Essentially, GAMP uses a high-dimensional approximation of loopy belief propagation to convert the MMSE or MAP inference problems into a sequence of tractable scalar inference problems. 

GAMP is well motivated in the case that $\vec{A}$ is a realization of a large random matrix with i.i.d zero-mean sub-Gaussian entries.
For such $\vec{A}$, in the large-system limit (i.e., $M,N\rightarrow\infty$ for fixed $M/N\in\Real_+$), GAMP is characterized by a state evolution whose fixed points, when unique, are MMSE or MAP optimal \cite{Rangan:ISIT:11,Bayati:ISIT:12,Javanmard:II:13}.
Furthermore, for \emph{generic} $\vec{A}$, it has been shown \cite{Rangan:ISIT:13} that MAP-GAMP's fixed points coincide with the critical points of the cost function \eqref{Jmap} and that MMSE-GAMP's fixed points coincide with those of a Bethe free entropy \cite{Krzakala:ISIT:14}, as discussed in detail in \secref{Bethe}.

For generic $\vec{A}$, however, GAMP may not reach its fixed points, i.e., it may diverge (e.g., \cite{Caltagirone:ISIT:14}).
The convergence of GAMP has been fully characterized in \cite{Rangan:ISIT:14} for the simple case that $p_{\X_n}$ and $p_{\Y_m|\Z_m}$ are Gaussian.
There, it was shown that Gaussian-GAMP converges if and only if the peak-to-average ratio of the squared singular values of $\vec{A}$ is sufficiently small.  
A \emph{damping} modification was then proposed in \cite{Rangan:ISIT:14} that guarantees the convergence of Gaussian-GAMP with arbitrary $\vec{A}$, at the expense of a slower convergence rate.
For strictly log-concave $p_{\X_n}$ and $p_{\Y_m|\Z_m}$, the \emph{local} convergence of GAMP was also characterized in \cite{Rangan:ISIT:14}.
However, the global convergence of GAMP under generic $\vec{A}$, $p_{\X_n}$, and $p_{\Y_m|\Z_m}$ is not yet understood. 

Because of its practical importance, prior work has attempted to robustify the convergence of GAMP in the face of ``difficult'' $\vec{A}$ (e.g., high peak-to-average singular values) for generic $p_{\X_n}$ and $p_{\Y_m|\Z_m}$.
For example, ``swept'' GAMP (SwAMP) \cite{Manoel:swamp:14} updates the estimates of $\{\X_n\}_{n=1}^N$ and $\{\Z_m\}_{m=1}^M$ sequentially, in contrast to GAMP, which updates them in parallel. 
Relative to GAMP, experiments in \cite{Manoel:swamp:14} show that SwAMP is much more robust to difficult $\vec{A}$, but it is slower and cannot facilitate fast implementations of $\vec{A}$ like an FFT.
As another example, the public-domain GAMPmatlab implementation \cite{GAMPmatlab} has included ``adaptive damping'' and ``mean removal'' mechanisms for some time, but they have never been described in the literature.

In this paper, we detail the most recent versions of GAMPmatlab's adaptive damping and mean-removal mechanisms, and we experimentally characterize their performance on non-zero-mean, rank-deficient, column-correlated, and ill-conditioned $\vec{A}$ matrices.
Our results show improved robustness relative to SwAMP and enhanced convergence speed.

%\emph{Notation}:
%  For matrices, we use boldface capital letters like $\vec{A}$, and we use $\vec{A}\tran$ to denote its transpose.
%  For vectors, we use boldface small letters like $\vec{x}$, and we use $\norm{\vec{x}}_p=(\sum_n |x_n|^p)^{1/p}$ to denote the $\ell_p$ norm, with $x_n=[\vec{x}]_n$ representing the $n^{th}$ element of $\vec{x}$.
%  Deterministic quantities are denoted using serif typeface (e.g., $x,\vec{x}$), while random quantities are denoted using san-serif typeface (e.g., $\X,\vX$).
%  For random variable\ $\X$, we write the pdf as $f_{\X}(x)$, the expectation as $\E\{\X\}$, and the variance as $\var\{\X\}$.
%  For a Gaussian random vector $\vX$ with mean $\vec{m}$ and covariance $\vec{C}$, we write the pdf as $\mc{N}(x;\vec{m},\vec{C})$.
% Finally, we use $\Real$ for the real field.

%%%%%%%%%%%%%%%%%%%%%%%%%%%%%%%%%%%%%%%%%%%%%%%%%%%%%%%%%%%%%%%%%%%%%%%%%%%%%
\section{Adaptively Damped GAMP} \label{sec:ADGAMP}

Damping is commonly used in loopy belief propagation to ``slow down'' the updates in an effort to promote convergence.
(See, e.g., \cite{Heskes:NIPS:02} for damping applied to the sum-product algorithm and \cite{Rangan:ISIT:14,Schniter:ALL:12,GAMPmatlab} for damping applied to GAMP.)
However, since not enough damping allows divergence while too much damping unnecessarily slows convergence, we are motivated to develop an \emph{adaptive} damping scheme that applies just the right amount of damping at each iteration.

\tabref{adgamp} details the proposed \emph{adaptively damped GAMP} (AD-GAMP) algorithm. 
Lines (R3)-(R6) and (R10) use an iteration-$t$-dependent damping parameter $\beta(t)\in(0,1]$ to slow the updates,\footnote{The GAMPmatlab implementation \cite{GAMPmatlab} allows one to disable damping in (R6) and/or (R10).}
and lines (R12)-(R18) adapt the parameter $\beta(t)$.
When $\beta(t)=1~\forall t$, AD-GAMP reduces to the original GAMP from \cite{Rangan:ISIT:11}.
Due to lack of space, we refer readers to \cite{Rangan:ISIT:11,Rangan:ISIT:13} for further details on GAMP.

\putTable{adgamp}{The adaptively damped GAMP algorithm.  In lines (R1) and (R8),
$g'_{\Z_m}$ and $g'_{\X_n}$ denote the derivatives of $g_{\Z_m}$ and $g_{\X_n}$ w.r.t their first arguments.}
{\footnotesize
\[\arraycolsep=0.7pt
\begin{array}{|l@{}rcl@{}r|}\hline 
% %%%%%%%%%%% DEFNS %%%%%%%%%%%%%%%%%%%%%%%%
 \multicolumn{4}{|l}{\textsf{definitions for MMSE-GAMP:}}&\\
 &g_{\Z_m}(\hat{p},\nu^p)
  &\defn& \int z \,f_{\Z_m\!}(z;\hat{p},\nu^p) dz &\text{\scriptsize(D1)}\\
  &&& \text{for~} f_{\Z_m\!}(z;\hat{p},\nu^p) 
      \defn \frac{p_{\Y_m|\Z_m\!}(y_m|z) \mc{N}(z;\hat{p},\nu^p)}{B_m(\hat{p},\nu^p)}&\\
  &&& \text{and~} B_m(\hat{p},\nu^p) 
      \defn \int p_{\Y_m|\Z_m\!}(y_m|z) \,\mc{N}(z;\hat{p},\nu^p) dz&\\
 &g_{\X_n\!}(\hat{r},\nu^r)
  &\defn& \int x \,f_{\X_n\!}(x;\hat{r},\nu^r) dx &\text{\scriptsize(D2)}\\
  &&& \text{for~} f_{\X_n\!}(x;\hat{r},\nu^r) 
      \defn \frac{p_{\X_n\!}(x) \mc{N}(x;\hat{r},\nu^r)}{C_n(\hat{r},\nu^r)}&\\
  &&& \text{and~} C_n(\hat{r},\nu^r) \defn \int p_{\X_n\!}(x) \,\mc{N}(x;\hat{r},\nu^r) dx&\\[1mm]
 \multicolumn{4}{|l}{\textsf{definitions for MAP-GAMP:}}&\\
 &g_{\Z_m\!}(\hat{p},\nu^p)
  &\defn& \argmax_{z} \ln p_{\Y_m|\Z_m\!}(y_m|z) + \frac{1}{2\nu^p}|z-\hat{p}|^2 &\text{\scriptsize(D3)}\\
 &g_{\X_n\!}(\hat{r},\nu^r)
  &\defn& \argmax_{x} \ln p_{\X_n\!}(x) + \frac{1}{2\nu^r}|x-\hat{r}|^2 &\text{\scriptsize(D4)}\\[1mm]
  \hline
% %%%%%%%%%%% INPUTS %%%%%%%%%%%%%%%%%%%%%%%
  \multicolumn{2}{|l}{\textsf{inputs:}}&&&\\
  &\multicolumn{4}{l|}{\quad \forall m,n\!: g_{\Z_m\!}, g_{\X_n\!}, \hat{x}_n(1), \nu^x_n(1), a_{mn}, T_{\max}\geq 1, \epsilon\geq 0}\\ 
  &\multicolumn{4}{l|}{\quad T_{\beta}\geq 0, \beta_{\max}\in(0,1], \beta_{\min}\in[0,\beta_{\max}], G_{\text{pass}}\geq 1, G_{\text{fail}}<1 }\\[1mm]
  %%%%%%%% INITS %%%%%%%%%%%%%%%%%%%%%%%
  \multicolumn{2}{|l}{\textsf{initialize:}}&&&\\
%  &\forall n\!:\! 
%   \hat{x}_{n}(1) &=& \int_{x} x\, p_{\X_n}\!(x), 
%   \nu^x_n(1) \!=\! \int_{x} |x-\hat{x}_n(1)|^2  p_{\X_n}\!(x) &\text{\scriptsize(I1)}\\
  &\forall m\!: 
   \nu^p_m(1) &=&\textstyle \sum_{n\!=\!1}^{N} |a_{mn}|^2 \nu^x_{n}(1),~~
   \hat{p}_m(1) \!=\! \sum_{n\!=\!1}^{N} a_{mn} \hat{x}_{n}(1) &\text{\scriptsize(I2)}\\
  &J(1) &=& \infty, ~\beta(1) = 1, ~t = 1 &\text{\scriptsize(I3)}\\[1mm]
  %%%%%%%%%%%%%%%% Begin %%%%%%%%%%%%%%%%%%%%%%%%%%%%%
  \multicolumn{3}{|l}{\textsf{while $t \leq T_{\max}$,}}&&\\
  &\forall m\!: 
   \nu^z_m(t)
   &=& \nu^p_m(t)\, g'_{\Z_m\!}(\hat{p}_m(t),\nu^p_m(t)) &\text{\scriptsize(R1)}\\
  &\forall m\!: 
   \hat{z}_m(t)
   &=& g_{\Z_m\!}(\hat{p}_m(t),\nu^p_m(t)) &\text{\scriptsize(R2)}\\
  &\forall m\!: 
   \nu^s_m(t)
   &=& \beta(t)\Big(1\!-\!\frac{\nu^z_m(t)}{\nu^p_m(t)}\Big)\frac{1}{\nu^p_m(t)} \!+\! \big(1\!-\!\beta(t)\big)\nu^s_m(t\!-\!1) &\text{\scriptsize(R3)}\\
  &\forall m\!: 
   \hat{s}_m(t)
   &=& \beta(t)\frac{\hat{z}_m(t)-\hat{p}_m(t)}{\nu^p_m(t)} \!+\! \big(1\!-\!\beta(t)\big)\hat{s}_m(t\!-\!1)&\text{\scriptsize(R4)}\\
  &\forall n\!: 
   \tilde{x}_n(t)
   &=&  \beta(t)\hat{x}_n(t) + \big(1\!-\!\beta(t)\big)\tilde{x}_n(t\!-\!1) &\text{\scriptsize(R5)}\\
  &\forall n\!: 
   \nu^r_n(t)
   &=& \beta(t)\frac{1}{\sum_{m=1}^{M} |a_{mn}|^2 \nu^s_m(t)} \!+\! \big(1\!-\!\beta(t)\big)\nu^r_n(t\!-\!1) 
        &\text{\scriptsize(R6)}\\
  &\forall n\!: 
   \hat{r}_n(t)
   &=& \textstyle \tilde{x}_n(t)+ \nu^r_n(t) \sum_{m=1}^{M} \!a_{mn}\herm
        \hat{s}_{m}(t)  &\text{\scriptsize(R7)}\\
  &\forall n\!: 
   \nu^x_n(t\!+\!\!1)
   &=& \nu^r_n(t)\, g'_{\X_n\!}(\hat{r}_n(t),\nu^r_n(t)) &\text{\scriptsize(R8)}\\
  &\forall n\!: 
   \hat{x}_{n}(t\!+\!\!1)
   &=& g_{\X_n\!}(\hat{r}_n(t),\nu^r_n(t)) &\text{\scriptsize(R9)}\\
  &\forall m\!: 
 \nu^p_m(t\!+\!\!1)
   &=& \beta(t) \textstyle \sum_{n=1}^{N} |a_{mn}|^2 \nu^x_{n}(t\!+\!1) + (1\!-\!\beta(t)\big) \nu^p_m(t) &\text{\scriptsize(R10)} \\
  &\forall m\!: 
   \hat{p}_m(t\!+\!\!1)
   &=& \sum_{n=1}^{N} a_{mn} \hat{x}_{n}(t\!+\!1) - \nu^p_m(t\!+\!1) \,\hat{s}_m(t) &\text{\scriptsize(R11)}\\
%&\forall m\!: 
%  \tilde{p}_m(t\!+\!\!1) 
%  &=& \textsf{FP}\!\big([\ovec{A}\hvec{x}(t\!+\!\!1)]_m,\hat{p}_m(t\!+\!\!1),\nu^p_m(t\!+\!\!1), p_{\Y_m|\Z_m}\!\big)\! &\text{\scriptsize(R12)}\\
%&Q(t\!+\!\!1) &=& J(\tilde{\vec{p}}(t\!+\!\!1),\vec{\nu^p}(t\!+\!\!1),\hvec{r}(t\!+\!\!1),\vec{\nu^r}(t\!+\!\!1)) &\text{\scriptsize(R13)} \\    
&J(t\!+\!\!1) &=& \textsf{\scriptsize eqn \eqref{Jmap} for MAP-GAMP or eqn \eqref{Jmse} for MMSE-GAMP} &\text{\scriptsize(R12)} \\
      &\multicolumn{3}{l}{\hspace{5mm} \textsf{if~} \hspace{1.5 mm} J(t\!+\!1) \leq \max_{\tau=\max\{t-T_\beta,1\},...,t}J(\tau) \textsf{~or~} \beta(t) = \beta_{\min}} &\text{\scriptsize(R13)}\\
      &\multicolumn{3}{l}{\hspace{5mm} \textsf{then~}
      \hspace{1.5mm }\textsf{if} \hspace{3.5 mm}
        %\textsf{tol}\big(\hvec{x}(t), \hvec{x}(t\!+\!1)\big)
        \|\hvec{x}(t)-\hvec{x}(t\!+\!1)\|/\|\hvec{x}(t\!+\!1)\|
        < \epsilon,} &\text{\scriptsize(R14)}\\
      &\multicolumn{3}{l}{\hspace{13 mm} \textsf{then~} 
            \hspace{1.5mm} \textsf{stop}} &\text{\scriptsize(R15)} \\
      &\multicolumn{3}{l}{\hspace{13 mm} \textsf{else~} 
            \hspace{1.5mm} \beta(t\!+\!1) = \min\{\beta_{\max},G_{\text{pass}}\beta(t)\}} &\text{\scriptsize(R16)} \\
      &&& \hspace{8.25mm} t = t\!+\!1 &\text{\scriptsize(R17)} \\
      &\multicolumn{3}{l}{\hspace{5mm} \textsf{else~}  
            \hspace{1.5 mm} \beta(t) = \max\{\beta_{\min},G_{\text{fail}}\beta(t)\}, 
      } &\text{\scriptsize(R18)}\\
%  &\multicolumn{3}{l}{\hspace{3.5mm}
%      \textsf{if} \sum_{n=1}^N|\hat{x}_n(t\!+\!1)-\hat{x}_n(t)|^2 < \epsilon_{\textsf{gamp}} \sum_{n=1}^N|\hat{x}_n(t)|^2,
%      \textsf{break}} & \text{\scriptsize(R15)}\\
  \multicolumn{2}{|l}{\textsf{end}}&&&\\[1mm]
  \multicolumn{5}{|l|}{\textsf{outputs:~}
        \forall m,n\!:
        \hat{r}_n(t),\nu^r_n(t), 
        \hat{p}_m(t\!+\!1),\nu^p_m(t\!+\!1), 
        \hat{x}_n(t\!+\!1),\nu^x_n(t\!+\!1)
        }\\[1mm]
  \hline
\end{array}
\]
}

\subsection{Damping Adaptation} \label{sec:AD}

The damping adaptation mechanism in AD-GAMP works as follows.
Line (R12) computes the current cost $J(t\!+\!1)$, as described in the sequel.
Line (R13) then checks evaluates whether the current iteration ``passes'' or ``fails'': it passes if the current cost is at least as good as the worst cost over the last $T_\beta\geq 0$ iterations or if $\beta(t)$ is already at its minimum allowed value $\beta_{\min}$, else it fails.
If the iteration passes, (R14)-(R15) implement a stopping condition, (R16) increases $\beta(t)$ by a factor $G_{\text{pass}}\!\geq\!1$ (up to the maximum value $\beta_{\max}$), and (R17) increments the counter $t$.
If the iteration fails, (R18) decreases $\beta(t)$ by a factor $G_{\text{fail}}\!<\!1$ (down to the minimum value $\beta_{\min}$) and the counter $t$ is \emph{not} advanced, causing AD-GAMP to re-try the $t$th iteration with the new value of $\beta(t)$.

In the MAP case, line (R12) simply computes the cost $J(t\!+\!1)=J_\MAP(\hvec{x}(t\!+\!1))$ for $J_\MAP$ from \eqref{Jmap}.
The MMSE case, which is more involved, will be described next.

%%%%%%%%%%%%%%%%%%%%%%%%%%%%%%%%%%%%%%%%%%%%%%%%%%%%%%%%%%%%%%%%%%%%%%%%%%%%%
\subsection{MMSE-GAMP Cost Evaluation} \label{sec:Bethe}

As proven in \cite{Rangan:ISIT:13} and interpreted in the context of Bethe free entropy in \cite{Krzakala:ISIT:14}, 
the fixed points of MMSE-GAMP are critical points of the optimization problem 
\begin{eqnarray}
(f_{\vX},f_{\vZ}) 
&=& \argmin_{b_{\vX},b_{\vZ}} 
J_\bethe(b_{\vX},b_{\vZ}) \st\! \E\{\vZ|b_{\vZ}\}\!=\!\vec{A}\E\{\vX|b_{\vX}\}
        \quad\label{eq:MMSEopt} \\
J_\bethe(b_{\vX},b_{\vZ})
&\defn& D\big(b_{\vX} \| p_{\vX}\big) 
        + D\big(b_{\vZ} \| p_{\vY|\vZ}Z^{-1}\big) + H\big(b_{\vZ}, \vec{\nu}^p\big)
        \quad\label{eq:Bethe_cost} \\
H\big(b_{\vZ}, \vec{\nu}^p\big)
&\defn& \frac{1}{2}\sum_{m=1}^{M} \bigg( \frac{\var\{\Z_m|b_{\Z_m}\}}{\nu^p_m} + \ln 2\pi\nu^p_m \bigg)
        \label{eq:HG} ,
\end{eqnarray}
where $b_{\vX}$ and $b_{\vZ}$ are separable pdfs,
$Z^{-1}\!\defn\!\int p_{\vY|\vZ}(\vec{y}|\vec{z}) d\vec{z}$ is the scaling factor that renders $p_{\vY|\vZ}(\vec{y}|\vec{z})Z^{-1}$ a valid pdf over $\vec{z}\!\in\!\Real^M$,
$D(\cdot\|\cdot)$ denotes Kullback-Leibler (KL) divergence, and 
$H(b_{\vZ})$ is an upper bound on the entropy of $b_{\vZ}$ that is tight when $b_{\vZ}$ is independent Gaussian with variances in $\vec{\nu}^p$.
In other words, the pdfs 
$f_{\vX}(\vec{x};\hvec{r},\vec{\nu^r}) = \prod_{n=1}^{N} f_{\X_n}(x_n;\hat{r}_n,\nu^r_n)$ 
and 
$f_{\vZ}(\vec{z};\hvec{p},\vec{\nu^p}) = \prod_{m=1}^{M} f_{\Z_m}(z_m;\hat{p}_m,\nu^p_m)$ 
given in lines (D1) and (D2) of \tabref{adgamp} are critical points of \eqref{MMSEopt} for fixed-point versions of $\hvec{r}, \vec{\nu^r}, \hvec{p}, \vec{\nu^p}$.

Since $f_{\vX}$ and $f_{\vZ}$ are functions of $\hvec{r}, \vec{\nu^r}, \hvec{p}, \vec{\nu^p}$, the cost $J_\bethe$ can be written in terms of these quantities as well.
For this, we first note 
\begin{eqnarray}
  \lefteqn{ 
  D\big( f_{\X_n\!} \big\| p_{\X_n\!} \big)
  = \int \!\!f_{\X_n\!}(x;\hat{r}_n,\nu^r_n)
        \ln \frac{p_{\X_n\!}(x) \mc{N}(x;\hat{r}_n,\nu^r_n)}{p_{\X_n\!}(x)C_n(\hat{r}_n,\nu^r_n)} dx 
  } \\
  &=& -\ln C_n(\hat{r}_n,\nu^r_n) - \frac{\ln 2\pi\nu^r_n}{2}
        - \!\!\int \!\!f_{\X_n\!}(x;\hat{r}_n,\nu^r_n) \frac{|x-\hat{r}_n|^2}{2\nu^r_n} dx 
        \quad \, \\
  &=& -\ln C_n(\hat{r}_n,\nu^r_n) - \frac{\ln 2\pi\nu^r_n}{2}
        - \frac{|\hat{x}_n-\hat{r}_n|^2 + \nu^x_n}{2\nu^r_n} ,
        \label{eq:KLX}
\end{eqnarray}
where $\hat{x}_n$ and $\nu^x_n$ are the mean and variance of $f_{\X_n}(\cdot;\hat{r}_n,\nu^r_n)$ from (R9) and (R8).
Following a similar procedure, 
\begin{eqnarray}
 \lefteqn{
 D\big(f_{\Z_m\!}\|p_{\Y_m|\Z_m\!}Z_m^{-1}\big)  
 }\nonumber\\
 &=& -\ln \frac{B_m(\hat{p}_m,\nu^p_m)}{Z_m} - \frac{\ln 2\pi\nu_m^p}{2}
        - \frac{|\hat{z}_m\!-\!\hat{p}_m|^2 + \nu^z_m}{2\nu^p_m} ,
        \label{eq:KLZ}
\end{eqnarray}
where $\hat{z}_m$ and $\nu^z_m$ are the mean and variance of $f_{\Z_m}(\cdot;\hat{p}_m,\nu^p_m)$ from (R2) and (R1).
Then, since $D(f_{\vX}\|p_{\vX})=\sum_{n=1}^{N} D\big(f_{\X_n} \| p_{\X_n}\big)$ and
$D(f_{\vZ}\|p_{\vY|\vZ}Z^{-1})=\sum_{m=1}^{M} D\big(f_{\Z_m} \| p_{\Y_m|\Z_m} Z_m^{-1}\big)$, 
\eqref{Bethe_cost} and \eqref{HG} imply
\begin{eqnarray}
  \lefteqn{
  J_\bethe(\hvec{r},\vec{\nu}^r,\hvec{p},\vec{\nu}^p)  
  = -\sum_{m=1}^{M} \!\!\Bigg( \!\!\ln B_m(\hat{p}_m,\nu^p_m)
        + \frac{|\hat{z}_m\!-\!\hat{p}_m|^2}{2\nu^p_m} \Bigg) 
  }\nonumber\\
  &&-\sum_{n=1}^{N} \!\Bigg( \!\ln C_n(\hat{r}_n,\nu^r_n) + \frac{\ln \nu_n^r}{2}
        + \frac{\nu^x_n \!+\! |\hat{x}_n\!-\!\hat{r}_n|^2}{2\nu^r_n} \!\Bigg)
        \!+\! \textsf{const},
        \label{eq:Bethe_cost2}
  \hspace{5mm}
\end{eqnarray}
where we have written $J_\bethe(f_{\vX},f_{\vZ})$ as ``$J_\bethe(\hvec{r},\vec{\nu}^r,\hvec{p},\vec{\nu}^p)$'' to make the $(\hvec{r},\vec{\nu}^r,\hvec{p},\vec{\nu}^p)$-dependence clear, and where $\textsf{const}$ collects terms invariant to $(\hvec{r},\vec{\nu^r},\hvec{p},\vec{\nu^p})$.

%\begin{eqnarray}
% \lefteqn{
% D\big(f_{\vZ} \| p_{\vY|\vZ}Z^{-1}\big)  = {\textstyle \sum_{m=1}^{M} D\big(f_{\Z_m|\P_m} \| p_{\Y_m| \Z_m}Z^{-1}_m\big)} 
% }\\
% &=& -\sum_{m=1}^{M}\! \Bigg( \!\ln \frac{B_m(\hat{p}_m,\nu^p_m)}{Z_m(y_m)} + \frac{\ln 2\pi\nu_m^p}{2}
%       + \frac{\nu^z_m \!+\! |\hat{z}_m\!-\!\hat{p}_m|^2}{2\nu^p_m}\! \Bigg). \nonumber
%\label{eq:KLZ}
%\end{eqnarray}
%\begin{align}
%  &D\big(f_{\vX} \| p_{\vX}\big)  = {\textstyle \sum_{n=1}^{N} D\big(f_{\X_n} \| p_{\X_n}\big)}\\
%  &= -\sum_{n=1}^{N}\! \Bigg( \!\ln C_n(\hat{r}_n,\nu^r_n) + \frac{\ln 2\pi\nu_n^r}{2}
%       + \frac{\nu^x_n \!+\! |\hat{x}_n\!-\!\hat{r}_n|^2}{2\nu^r_n}\! \Bigg). \nonumber
%\end{align}
%where $\ln Z_m(\overline{y}_m)$ is constant w.r.t $\hat{p}_m$ and $\nu^p_m$ and can thus be ignored.
%We then insert \eqref{KLX} and \eqref{KLZ} into \eqref{Bethe_cost} and simplify, yielding
%\begin{align}
%  &J(\hvec{p},\vec{\nu^p},\hvec{r},\vec{\nu^r})  
%  = -\sum_{m=1}^{M} \!\Bigg( \!\ln \frac{B_m(\hat{p}_m,\nu^p_m)}{Z_m(y_m)}
%       + \frac{|\hat{z}_m\!-\!\hat{p}_m|^2}{2\nu^p_m} \Bigg) \nonumber \\
%  &-\sum_{n=1}^{N} \!\Bigg( \!\ln C_n(\hat{r}_n,\nu^r_n) + \frac{\ln 2\pi\nu_n^r}{2}
%       + \frac{\nu^x_n \!+\! |\hat{x}_n\!-\!\hat{r}_n|^2}{2\nu^r_n} \Bigg),
%       \label{eq:Bethe_cost2}
%\end{align}
%where we change the notation to emphasize that the cost \eqref{Bethe_cost2} depends only on the fixed points $\hvec{p},\vec{\nu^p},\hvec{r}$, and $\vec{\nu^r}$.

Note that the iteration-$t$ MMSE-GAMP cost is not obtained simply by plugging $(\hvec{r}(t), \vec{\nu^r}(t),\hvec{p}(t\!+\!1), \vec{\nu^p}(t\!+\!1))$ into \eqref{Bethe_cost2}, because the latter quantities do not necessarily yield $(f_{\vX},f_{\vZ})$ satisfying the moment-matching constraint $\E\{\vZ|f_{\vZ}\} \!=\! \vec{A}\E\{\vX|f_{\vX}\}$ from \eqref{MMSEopt}.
Thus, it was suggested in \cite{Krzakala:ISIT:14} to compute the cost as 
\begin{align}
J_\MSE(\hvec{r}(t), \vec{\nu^r}(t))
&= J_\bethe(\hvec{r}(t), \vec{\nu^r}(t), \tvec{p}, \vec{\nu^p}(t\!+\!1)), 
        \label{eq:Jmse}
\end{align}
for $\tvec{p}$ chosen to match the moment-matching constraint, i.e., for
\begin{equation}
  [\vec{A}\hvec{x}(t\!+\!1)]_m 
  = g_{\Z_m\!}\big(\tilde{p}_m,\nu^p_m(t\!+\!1)\big)
  \text{~for~$m=1,\dots,M$}
  \label{eq:fixed_p}
\end{equation}
where $\hat{x}_n(t\!+\!1) = g_{\X_n\!}\big(\hat{r}_n(t),\mu^r_n(t)\big)$ for $n=1,\dots,N$ from (R9).
Note that, since $\vec{\nu^p}(t\!+\!1)$ can be computed from $(\hvec{r}(t), \vec{\nu^r}(t))$ via (R8) and (R10), the left side of \eqref{Jmse} uses only $(\hvec{r}(t), \vec{\nu^r}(t))$.

In the case of an additive white Gaussian noise (AWGN), i.e., $p_{\Y_m|\Z_m\!}(y_m|z_m)=\mc{N}(z_m;y_m,\nu^w)$ with $\nu^w\!>\!0$, the function $g_{\Z_m\!}(\tilde{p}_m,\nu^p_m)$ is linear in $\tilde{p}_m$.
In this case, \cite{Krzakala:ISIT:14} showed that \eqref{fixed_p} can be solved in closed-form, yielding the solution 
\begin{equation}
  \tilde{p}_m
  = \big( (\nu^p_m(t\!+\!1)+\nu^w)[\vec{A}\hvec{x}(t\!+\!1)]_m - \nu^p_m(t\!+\!1)y_m \big)/\nu^w .
\end{equation}

For general $p_{\Y_m|\Z_m\!}$, however, the function $g_{\Z_m\!}(\tilde{p}_m,\nu^p_m)$ is non-linear in $\tilde{p}_m$ and difficult to invert in closed-form. 
Thus, we propose to solve \eqref{fixed_p} numerically using the regularized Newton's method detailed in \tabref{estInv}.
There, $\alpha\in(0,1]$ is a stepsize, $\phi \geq 0$ is a regularization parameter that keeps the update's denominator positive, and $I_{\max}$ is a maximum number of iterations, all of which should be tuned in accordance with $p_{\Y_m|\Z_m\!}$.
Meanwhile, $\tilde{p}_m(1)$ is an initialization that can be set at $\hat{p}_m(t\!+\!1)$ or $[\vec{A}\hvec{x}(t\!+\!1)]_m$ and $\epsilon_{\text{inv}}$ is a stopping tolerance.
Note that the functions $g_{\Z_m\!}$ and $g'_{\Z_m\!}$ employed in \tabref{estInv} are readily available from \tabref{adgamp}. 
%If an implementation of the second derivative $g''_{\Z_m\!}$ was readily available, then it would be more efficient to use pure Newton's method \cite[p.~88]{Bertsekas:Book:99}.

\putTable{estInv}{A regularized Newton's method to find the value of $\tilde{p}_m$ that solves $[\vec{A}\hvec{x}]_m=g_{\Z_m\!}(\tilde{p}_m,\nu^p_m)$ for a given $[\vec{A}\hvec{x}]_m$ and $\nu^p_m$.}
{\footnotesize
\[\arraycolsep=0.7pt
\begin{array}{|lrcl@{}r|}\hline 
 \multicolumn{5}{|l|}{\textsf{inputs:~~}}\\
 &\multicolumn{4}{l|}{\quad 
        %\forall m\!:
        g_{\Z_m\!},
        [\vec{A}\hvec{x}]_m,
        \nu^p_m,
        \tilde{p}_m(1),
        I_{\max}\geq 1, 
        \epsilon_{\text{inv}} \geq 0,
        \alpha \in \!(0,1],
        \phi \geq 0 
 }\\[1mm]
%       %%%%%%%%%%% DEFNS %%%%%%%%%%%%%%%%%%%%%%%%
% \multicolumn{2}{|l}{\textsf{definitions:}}&&&\\
% &g_{\Z_m\!}(\hat{p},\nu^p)
%  &\defn& \int z \,f_{\Z_m\!}(z;\hat{p},\nu^p) dz &\text{\scriptsize(D1)}\\
%  &&& \text{for~} f_{\Z_m\!}(z;\hat{p},\nu^p) 
%      \defn \frac{p_{\Y_m|\Z_m\!}(y_m|z) \mc{N}(z;\hat{p},\nu^p)}{B_m(\hat{p},\nu^p)}&\\
%  &&& \text{and~} B_m(\hat{p},\nu^p) 
%      \defn \int p_{\Y_m|\Z_m\!}(y_m|z) \,\mc{N}(z;\hat{p},\nu^p) dz&\\[1mm]
   %%%%%%%%%%%%%%%% Begin %%%%%%%%%%%%%%%%%%%%%%%%%%%%%
  \multicolumn{4}{|l}{\textsf{for $i=1:I_{\max}$,}}&\\
  &%\forall m\!: 
   e_m(i)
   &=& [\vec{A}\hvec{x}]_m - g_{\Z_m\!}\big(\tilde{p}_m(i),\nu^p_m\big) & \text{\scriptsize(F1)} \\
  &\multicolumn{3}{l}{\hspace{5.5mm}
      \textsf{if} \hspace{1.5 mm} \big|e_m(i) / g_{\Z_m\!}\big(\tilde{p}_m(i),\nu^p_m\big)\big| < \epsilon_{\text{inv}} ,
      ~\textsf{stop}} & \text{\scriptsize(F2)}\\[0.5mm]
  &%\forall m\!:
   \nabla_m(i) 
   &=& g_{\Z_m\!}'\big(\tilde{p}_m(i),\nu^p_m\big) & \text{\scriptsize(F3)}\\
  &%\forall m\!: 
   \tilde{p}_m(i\!+\!1) 
   &=& \tilde{p}_m(i) + \alpha \frac{e_m(i) \nabla_m(i)}{\nabla_m^2(i) + \phi} &\text{\scriptsize(F4)} \\
  \multicolumn{2}{|l}{\textsf{end}}&&&\\[1mm]
  \multicolumn{4}{|l}{\textsf{outputs:~~}
        %\forall m\!:
        \tilde{p}_m(i)
        }&\\[1mm]
  \hline
\end{array}
\]
}

%%%%%%%%%%%%%%%%%%%%%%%%%%%%%%%%%%%%%%%%%%%%%%%%%%%%%%%%%%%%%%%%%%%%%%%%%%%%%
\subsection{Mean Removal} \label{sec:MeanRem}

To mitigate the difficulties caused by $\vec{A}$ with non-zero mean entries, we propose to rewrite the linear system ``$\vec{z} = \vec{Ax}$'' in \eqref{z} as 
\begin{equation}
\underbrace{
\mat{ \vec{z}\\ z_{M+1}\\ z_{M+2} }
}_{\displaystyle \defn \ovec{z}} =
\underbrace{
\mat{ \tvec{A} &b_{12}\vec{\gamma} & b_{13}\vec{1}_M\\ 
b_{21}\vec{1}_N\herm &-b_{21}b_{12} &0 \\ 
b_{31}\vec{c}\herm &0 &-b_{31}b_{13} }
}_{\displaystyle \defn \ovec{A}}
\underbrace{\mat{\vec{x} \\ x_{N+1} \\ x_{N+2}}}_{\displaystyle \defn \ovec{x}} 
\label{eq:aug}
\end{equation}
where
$(\cdot)\herm$ is conjugate transpose, 
$\vec{1}_P \defn [1,\dots,1]\herm \in \Real^P$, and
\begin{align}
\mu &\defn \tfrac{1}{MN}\vec{1}_M\herm\vec{A}\vec{1}_N 
   \label{eq:mu}\\
\vec{\gamma} &\defn \tfrac{1}{N}\vec{A}\vec{1}_N 
   \label{eq:gamma}\\
\vec{c}\herm &\defn \tfrac{1}{M}\vec{1}_M\herm \big(\vec{A} - \mu\vec{1}_M\vec{1}_N\herm \big) 
   \label{eq:c}\\
\tvec{A} &\defn \vec{A} - \vec{\gamma} \vec{1}_N\herm - \vec{1}_M \vec{c}\herm .
   \label{eq:Atilde}
\end{align}
The advantage of \eqref{aug} is that the rows and columns of $\ovec{A}$ are approximately zero-mean. 
This can be seen by first verifying, via the definitions above, that 
$\vec{c}\herm\vec{1}_N=0$,
$\tvec{A}\vec{1}_N=\vec{0}$, 
and $\vec{1}_M\herm\tvec{A}=\vec{0}\herm$,
which implies that the elements in every row and column of $\tvec{A}$ are zero-mean.
Thus, for large $N$ and $M$, the elements in all but a vanishing fraction of the rows and columns in $\ovec{A}$ will also be zero-mean.
The mean-square coefficient size in the last two rows and columns of $\ovec{A}$ can be made to match that in $\tvec{A}$ via choice of $b_{12}, b_{13}, b_{21}, b_{31}$.

To understand the construction of \eqref{aug}, note that \eqref{Atilde} implies
\begin{align}
\vec{z} 
= \vec{Ax} 
= \tvec{A}\vec{x}
  + b_{12}\vec{\gamma} \underbrace{ \vec{1}_N\herm\vec{x}/b_{12} }_{\displaystyle \defn x_{N+1}}
  + b_{13}\vec{1}_M \underbrace{ \vec{c}\herm\vec{x}/b_{13} }_{\displaystyle \defn x_{N+2}} ,
\label{eq:decomp}
\end{align}
which explains the first $M$ rows of \eqref{aug}.
To satisfy the definitions in \eqref{decomp}, we then require that $z_{M+1}=0$ and $z_{M+2}=0$ in \eqref{aug}, which can be ensured through the Dirac-delta likelihood 
\begin{equation}
p_{\Y_m|\Z_m}(y_m|z_m) 
\defn  \delta(z_m)      \text{~~for~~} m \!\in\! \{M\!+\!1,M\!+\!2\}. 
\label{eq:DiracYZ}
\end{equation}
Meanwhile, we make no assumption about the newly added elements $x_{N+1}$ and $x_{N+2}$, and thus adopt the improper uniform prior
\begin{equation}
p_{\X_n}(x_n) 
\propto 1 \text{~~for~~} n \in \{N\!+\!1,N\!+\!2\}. 
\label{eq:NullX}
\end{equation}

In summary, the mean-removal approach suggested here runs GAMP or AD-GAMP (as in \tabref{adgamp}) with $\ovec{A}$ in place of $\vec{A}$ and with the likelihoods and priors augmented by \eqref{DiracYZ} and \eqref{NullX}.
It is important to note that, if multiplication by $\vec{A}$ and $\vec{A}\herm$ can be implemented using a fast transform (e.g., FFT), then multiplication by $\ovec{A}$ and $\ovec{A}\herm$ can too; for details, see the GAMPmatlab implementation \cite{GAMPmatlab}.

%%%%%%%%%%%%%%%%%%%%%%%%%%%%%%%%%%%%%%%%%%%%%%%%%%%%%%%%%%%%%%%%%%%%%%%%%%%%%
\section{Numerical Results}                                             \label{sec:results}

We numerically studied the recovery $\NMSE \defn \norm{\hvec{x} - \vec{x}}^2/\norm{\vec{x}}^2$ of 
SwAMP \cite{Manoel:swamp:14} and the MMSE version of the original GAMP from \cite{Rangan:ISIT:11} 
relative to the proposed mean-removed (M-GAMP) and adaptively damped (AD-GAMP) modifications, as well as their combination (MAD-GAMP).
In all experiments, the signal $\vec{x}$ was drawn Bernoulli-Gaussian (BG) 
with sparsity rate $\tau$ and length $N\!=\!1000$,
and performance was averaged over $100$ realizations.
Average $\NMSE$ was clipped to $0$~dB for plotting purposes.
The matrix $\vec{A}$ was drawn in one of four ways: 
\begin{enumerate}[leftmargin=4ex,itemsep=0mm]
\item[(a)] 
\textbf{Non-zero mean}: i.i.d $\A_{mn} \sim \mc{N}(\mu, \tfrac{1}{N})$ for a specified $\mu\neq 0$.  
\item[(b)] 
\textbf{Low-rank product}: $\mA \!=\! \tfrac{1}{N}\mU\mV$ with $\mU \!\in\! \Real^{M \times R}$, $\mV \!\in\! \Real^{R \times N}$, and i.i.d $\U_{mr},\V_{rn}\!\sim\!\mc{N}(0,1)$, for a specified $R$.
Note $\mA$ is rank deficient when $R \!<\! \min\{M,N\}$.  
\item[(c)]
\textbf{Column-correlated}: the rows of $\mA$ are independent zero-mean stationary Gauss-Markov processes with a specified correlation coefficient $\rho=\E\{\A_{mn}\A_{m,n+1}\herm\}/\E\{|\A_{mn}|^2\}$.
\item[(d)]
\textbf{Ill-conditioned}:
$\mA=\mU\vec{\Sigma}\mV\herm$ where $\mU$ and $\mV\herm$ are the left and right singular vector matrices of an i.i.d $\mc{N}(0,1)$ matrix and $\vec{\Sigma}$ is a singular value matrix such that $[\vec{\Sigma}]_{i,i}/[\vec{\Sigma}]_{i+1,i+1}=(\kappa)^{1/\min\{M,N\}}$ for $i=1,\dots,\min\{M,N\}\!-\!1$, with a specified condition number $\kappa>1$.
\end{enumerate}
For all algorithms, we used
$T_{\max} \!=\! 1000$
and
$\epsilon \!=\! 10^{-5}$. 
Unless otherwise noted, for adaptive damping, we used
$T_\beta \!=\! 0$,
$G_{\text{pass}} \!=\! 1.1$,
$G_{\text{fail}} \!=\! 0.5$,
$\beta_{\max} \!=\! 1$,
and
$\beta_{\min} \!=\! 0.01$.
For SwAMP, we used the authors' publicly available code \cite{SwAMP}.

First we experiment with compressive sensing (CS) in AWGN at
$\SNR \!\defn\! \E\{\norm{\vZ}^2\}/\E\{\norm{\vY-\vZ}^2\} \!=\! 60$ dB. 
For this, we used $M\!=\!500\!=\!N/2$ measurements and sparsity rate $\tau \!=\! 0.2$. 
As a reference, we compute a lower-bound on the achievable $\NMSE$ using a genie who knows the support of $\vec{x}$.
For non-zero-mean matrices, \figref{AWGN4x4}(a) shows that the proposed M-GAMP and MAD-GAMP provided near-genie performance for all tested means $\mu$.  
In contrast, GAMP only worked with zero-mean $\vec{A}$ and SwAMP with small-mean $\vec{A}$.
For low-rank product, correlated, and ill-conditioned matrices, \figsref{AWGN4x4}(b)-(d) show that AD-GAMP is slightly more robust than SwAMP and significantly more robust than GAMP.

\putFrag{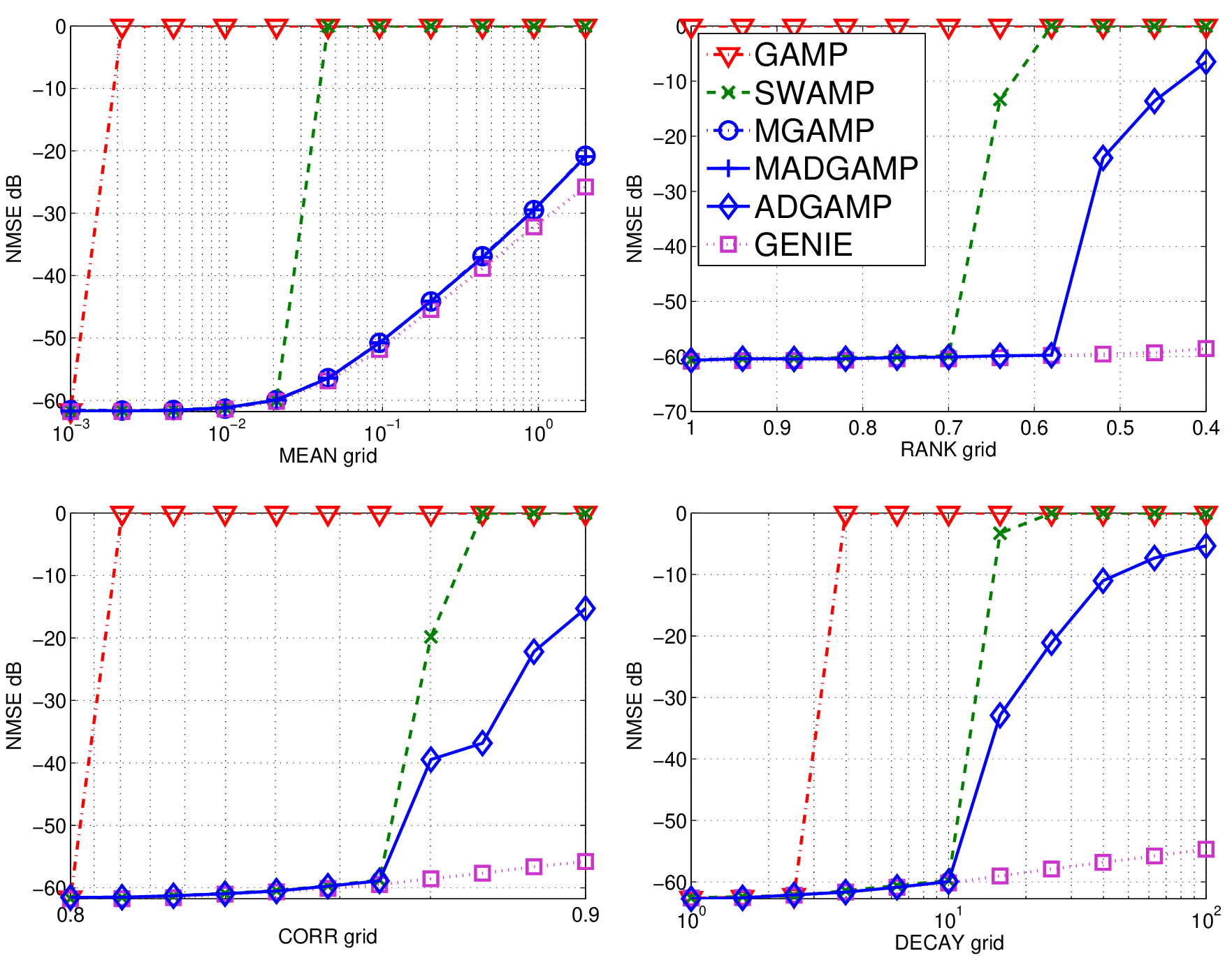}
        {AWGN compressive sensing under (a) non-zero-mean, (b) low-rank product, (c) column-correlated, and (d) ill-conditioned $\vec{A}$. }
        {\figsize}
        {\newcommand{\sz}{0.8}
         \psfrag{tau}[t][t[\sz]{$\rho$}
         \psfrag{NMSE dB}[][][\sz]{\sf NMSE [dB]}
         \newcommand{\szz}{0.6}
         \psfrag{RANK grid}[t][][\sz]{\sf (b) Rank Ratio $R/N$}
         \psfrag{MEAN grid}[t][][\sz]{\sf (a) Mean $\mu$}
         \psfrag{CORR grid}[t][][\sz]{\sf (c) Correlation $\rho$}
         \psfrag{DECAY grid}[t][][\sz]{\sf (d) Condition number $\kappa$}
         \psfrag{RANK}[][][\sz]{}
         \psfrag{MEAN}[][][\sz]{}
         \psfrag{CORR}[][][\sz]{}
         \psfrag{DECAY}[][][\sz]{}
         \psfrag{MADGAMP}[l][l][\szz]{\sf MAD-GAMP}
         \psfrag{MGAMP}[l][l][\szz]{\sf M-GAMP}
         \psfrag{ADGAMP}[l][l][\szz]{\sf AD-GAMP}
         \psfrag{SWAMP}[l][l][\szz]{\sf SwAMP}
         \psfrag{GENIE}[l][l][\szz]{\sf genie}
         \psfrag{GAMP}[l][l][\szz]{\sf GAMP}}

Next, we tried ``robust'' CS by repeating the previous experiment with sparsity rate $\tau\!=\!0.15$ and with 10\% of the observations (selected uniformly at random) replaced by ``outliers'' corrupted by AWGN at $\SNR\!=\!0$~dB.
For (M)AD-GAMP, we set $\beta_{\max}\!=\!0.1$ and $T_{\max}\!=\!2000$.
With non-zero-mean $\vec{A}$, \figref{AWGMN4x4}(a) shows increasing performance as we move from GAMP to M-GAMP to SwAMP to MAD-GAMP.
For low-rank product, correlated, and ill-conditioned matrices, \figref{AWGMN4x4}(b)-(d) show that SwAMP was slightly more robust than AD-GAMP, and both where much more robust than GAMP.
         
\putFrag{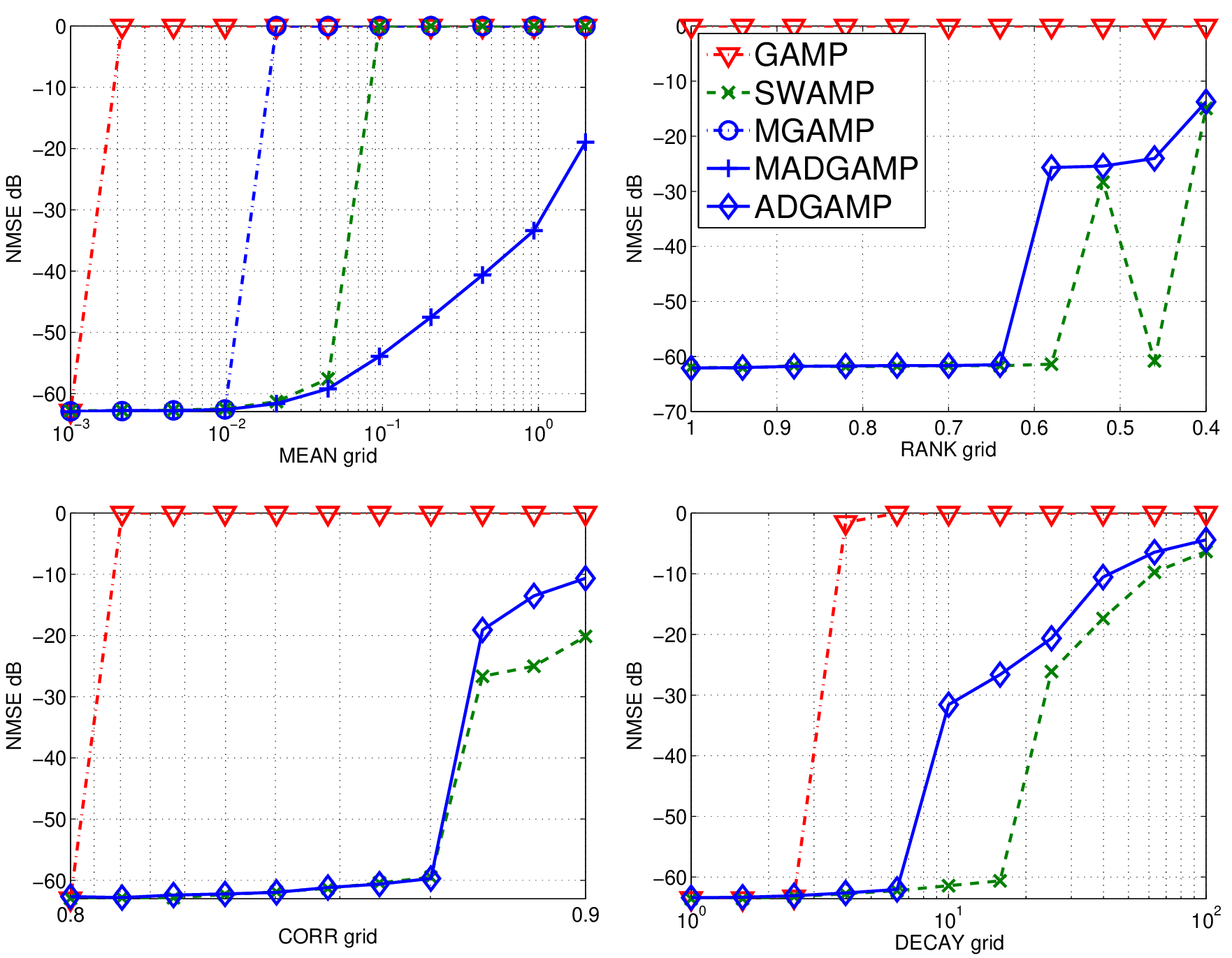}
        {``Robust'' compressive sensing under (a) non-zero-mean, (b) low-rank product, (c) column-correlated, and (d) ill-conditioned $\vec{A}$.}
        {\figsize}
        {\newcommand{\sz}{0.8}
         \psfrag{tau}[t][t[\sz]{$\rho$}
         \psfrag{NMSE dB}[][][\sz]{\sf NMSE [dB]}
         \newcommand{\szz}{0.6}
         \psfrag{RANK grid}[t][][\sz]{\sf (b) Rank Ratio $R/N$}
         \psfrag{MEAN grid}[t][][\sz]{\sf (a) Mean $\mu$}
         \psfrag{CORR grid}[t][][\sz]{\sf (c) Correlation $\rho$}
         \psfrag{DECAY grid}[t][][\sz]{\sf (d) Condition number $\kappa$}
         \psfrag{RANK}[][][\sz]{}
         \psfrag{MEAN}[][][\sz]{}
         \psfrag{CORR}[][][\sz]{}
         \psfrag{DECAY}[][][\sz]{}
         \psfrag{MADGAMP}[l][l][\szz]{\sf MAD-GAMP}
         \psfrag{MGAMP}[l][l][\szz]{\sf M-GAMP}
         \psfrag{ADGAMP}[l][l][\szz]{\sf AD-GAMP}
         \psfrag{SWAMP}[l][l][\szz]{\sf SwAMP}
         \psfrag{GENIE}[l][l][\szz]{\sf genie}
         \psfrag{GAMP}[l][l][\szz]{\sf GAMP}}
         
Finally, we experimented with noiseless $1$-bit CS \cite{Kamilov:TSP:12}, where $\vec{y} \!=\! \sgn(\vec{Ax})$, using $M\!=\!3000$ measurements and sparsity ratio $\tau\!=\!0.125$. 
In each realization, the empirical mean was subtracted from the non-zero entries of $\vec{x}$ to prevent $y_m\!=\!1~\forall m$. 
For (M)AD-GAMP, we used $\beta_{\max} \!=\! 0.5$.
For SwAMP, we increased the stopping tolerance to $\epsilon = 5\times 10^{-5}$, as it significantly improved runtime without degrading accuracy.
For non-zero-mean $\vec{A}$, \figref{PROBIT4x4}(a) shows that M-GAMP and MAD-GAMP were more robust than SwAMP, which was in turn much more robust than GAMP.
For low-rank product, correlated, and ill-conditioned matrices, \figsref{PROBIT4x4}(b)-(d) show that MAD-GAMP and SwAMP gave similarly robust performance, while the original GAMP was very fragile.
 
\putFrag{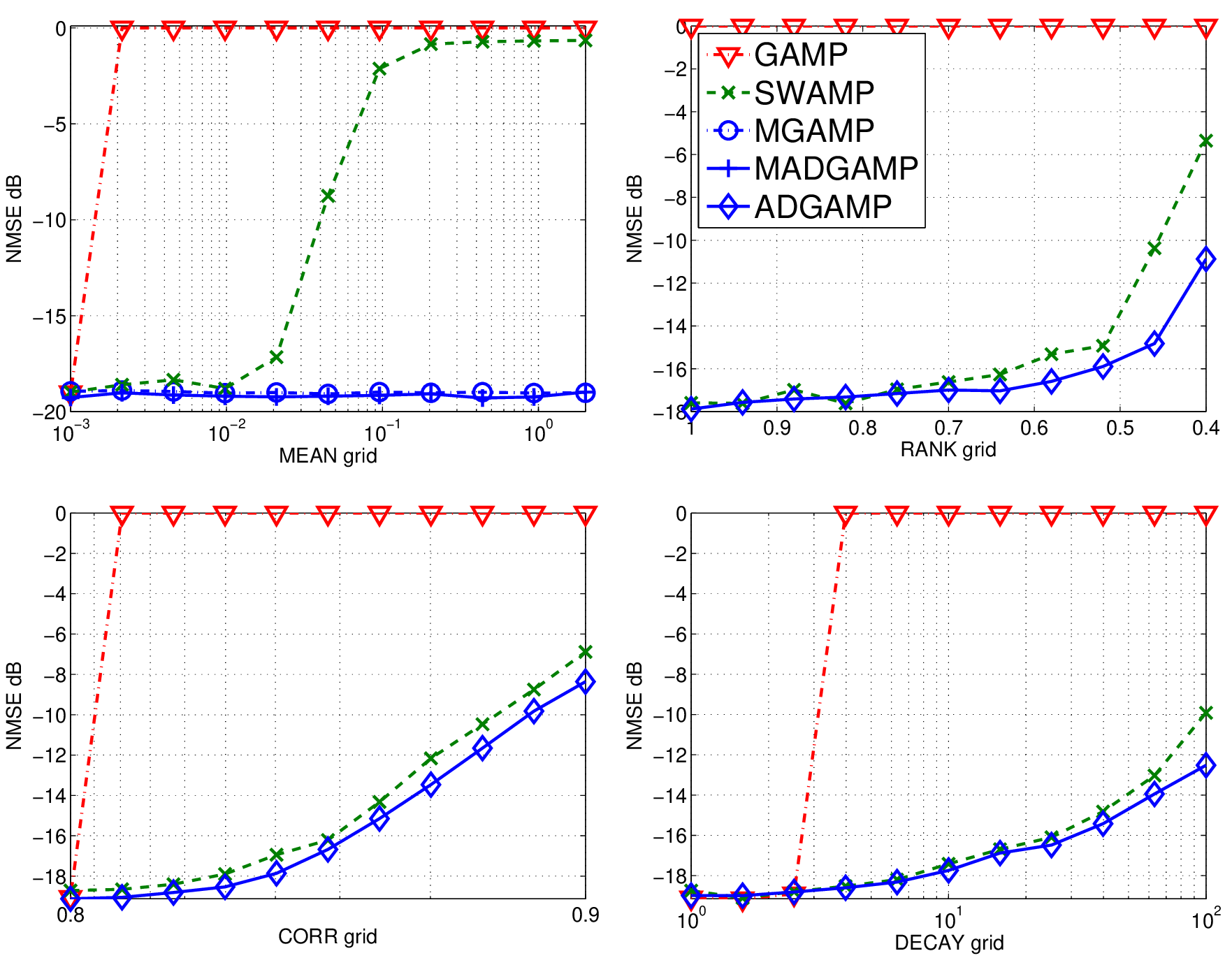}
        {$1$-bit compressive sensing under (a) non-zero-mean, (b) low-rank product, (c) column-correlated, and (d) ill-conditioned $\vec{A}$. }
        {\figsize}
        {\newcommand{\sz}{0.8}
         \psfrag{tau}[t][t[\sz]{$\rho$}
         \psfrag{NMSE dB}[][][\sz]{\sf NMSE [dB]}
         \newcommand{\szz}{0.6}
         \psfrag{RANK grid}[t][][\sz]{\sf (b) Rank Ratio $R/N$}
         \psfrag{MEAN grid}[t][][\sz]{\sf (a) Mean $\mu$}
         \psfrag{CORR grid}[t][][\sz]{\sf (c) Correlation $\rho$}
         \psfrag{DECAY grid}[t][][\sz]{\sf (d) Condition number $\kappa$}
         \psfrag{RANK}[][][\sz]{}
         \psfrag{MEAN}[][][\sz]{}
         \psfrag{CORR}[][][\sz]{}
         \psfrag{DECAY}[][][\sz]{}
         \psfrag{AMP}[l][l][\szz]{\sf MAD-GAMP}
         \psfrag{MADGAMP}[l][l][\szz]{\sf MAD-GAMP}
         \psfrag{ADGAMP}[l][l][\szz]{\sf AD-GAMP}
         \psfrag{MGAMP}[l][l][\szz]{\sf M-GAMP}
         \psfrag{SWAMP}[l][l][\szz]{\sf SwAMP}
         \psfrag{GAMP}[l][l][\szz]{\sf GAMP}}
         
Finally, we compare the convergence speed of MAD-GAMP to SwAMP.
For each problem, we chose a setting that allowed MAD-GAMP and SwAMP to converge for each matrix type. 
\tabref{runtime} shows that, on the whole, MAD-GAMP ran several times faster than SwAMP but used more iterations.
Thus, it may be possible to reduce SwAMP's runtime to below that of MAD-GAMP using a more efficient (e.g., BLAS-based) implementation, at least for explicit $\vec{A}$.
%At smaller values of $N$, though, much of MAD-GAMP's runtime is overhead associated with GAMPmatlab's object-oriented implementation.  
%For example, if we increase $N$ 
%from $1000$ to $10\,000$ in the AWGN experiment, then the (MAD-GAMP,\,SwAMP) runtimes go from $(0.45,0.94)$ to $(38,1530)$ seconds.
%the speedup goes from roughly $2\times$ to two orders-of-magnitude, as shown by the last row of \tabref{runtime}.
When $\vec{A}$ has a fast $O(N\log N)$ implementation (e.g., FFT), only (M)AD-GAMP will be able to exploit the reduced complexity.

%%Old table 10-24-14
%\putTable{runtime}{Average runtime (in seconds) of MAD-GAMP and SwAMP for various combinations of problem type and matrix type.  
%%The top 3 rows used $N=1000$ and the bottom row used $N=5000$.
%}
%{
%\vspace{2 mm}
%\scriptsize
%\bgroup
%\def\arraystretch{1.1}
%\begin{tabular}{l@{\,}|@{\,}c@{\,}|@{\,}c@{\,}|@{\,}c@{\,}|@{\,}c@{\,}|@{\,}c@{\,}|@{\,}c@{\,}|@{\,}c@{\,}|@{\,}c@{\,}|}
%\cline{2-9} \multicolumn{1}{l@{\,}|}{} &\multicolumn{2}{c|@{\,}}{$\mu = 0.021$} &\multicolumn{2}{c|@{\,}}{$R/N = 0.64$}  
% &\multicolumn{2}{c|@{\,}}{$\rho = 0.8$}  &\multicolumn{2}{c|}{$\log_{10}\kappa = 1$}\\
%\cline{2-9}  \multicolumn{1}{l@{\,}|}{}
%&{\tiny \, MAD-GAMP} &{\tiny SwAMP} &{\tiny AD-GAMP} &{\tiny SwAMP} 
%        &{\tiny AD-GAMP} &{\tiny SwAMP} &{\tiny AD-GAMP} &{\tiny SwAMP}\\ \hline
%\multicolumn{1}{|@{\,}l@{\,}|}{AWGN} &\textbf{0.45} &0.94 &\textbf{0.91} &3.84 &\textbf{1.32} &4.20 &\textbf{0.81} &1.42   \\ \hline
%\multicolumn{1}{|@{\,}l@{\,}|}{$1$-bit} &\textbf{45.38} &387.53 &\textbf{42.93} &279.49 &\textbf{44.50} &285.98 &\textbf{38.89} &261.77 \\ \hline
%\multicolumn{1}{|@{\,}l@{\,}|}{Robust} &\textbf{2.34} &8.73 &\textbf{3.41} &9.94 &\textbf{4.49} &16.61 &\textbf{2.50} &15.73 \\ \hline
%
%%N = 5000
%%\hline
%%\multicolumn{1}{|@{\,}l@{\,}|}{AWGN} &\textbf{8.40} &290.31 &\textbf{16.24} &52.02 &\textbf{24.13} &86.93 &\textbf{15.53} &35.89   \\ \hline
%
%%N = 10000
%%\hline
%%\multicolumn{1}{|@{\,}l@{\,}|}{AWGN} &\textbf{37.69} &1530.89 &\textbf{63.56} &212.39 &\textbf{107.22} &357.51 &\textbf{61.43} &141.12   \\ \hline
%
%\end{tabular}
%\egroup
%\vspace{2 mm}
%}

%New table 11-18-14

\putTable{runtime}{Average runtime (in seconds) and \# iterations of MAD-GAMP and SwAMP for various problem types and matrix types.} 
{
\vspace{1 mm}
\scriptsize
\bgroup
\def\arraystretch{1.1}
\begin{tabular}{|@{\,}c@{\,}|l@{\,}||@{\,}c@{\,}|@{\,}c@{\,}||@{\,}c@{\,}|@{\,}c@{\,}||@{\,}c@{\,}|@{\,}c@{\,}||@{\,}c@{\,}|@{\,}c@{\,}|}
\cline{3-10} \multicolumn{2}{l@{\,}||}{} &\multicolumn{2}{c||@{\,}}{$\mu = 0.021$} &\multicolumn{2}{c||@{\,}}{$R/N = 0.64$}  
 &\multicolumn{2}{c||@{\,}}{$\rho = 0.8$}  &\multicolumn{2}{c|}{$\log_{10}\kappa = 1$}\\
\cline{3-10}  \multicolumn{2}{l@{\,}||}{}
&{\tiny \, MAD-GAMP} &{\tiny SwAMP} &{\tiny AD-GAMP} &{\tiny SwAMP} 
        &{\tiny AD-GAMP} &{\tiny SwAMP} &{\tiny AD-GAMP} &{\tiny SwAMP}\\ \hline
\multirow{3}{*}{\begin{sideways}{seconds}\end{sideways}}
&\multicolumn{1}{|@{\,}l@{\,}||}{AWGN} 
&\textbf{1.06} &1.90 &\textbf{0.88} &2.74 
&\textbf{1.36} &3.84 &\textbf{0.81} &1.49   \\ \cline{2-10}
&\multicolumn{1}{|@{\,}l@{\,}||}{$1$-bit} 
&\textbf{53.34} &83.21 &\textbf{49.22} &137.46 
&\textbf{42.32} &149.40 &\textbf{50.25} &117.62 \\ \cline{2-10}
&\multicolumn{1}{|@{\,}l@{\,}||}{Robust} 
&\textbf{3.47} &8.81 &\textbf{2.66} &11.13
&\textbf{3.33} &15.70 &\textbf{2.38} &12.22 \\ \hline \hline

\multirow{3}{*}{\begin{sideways}{\# iters}\end{sideways}}
&\multicolumn{1}{|@{\,}l@{\,}||}{AWGN} 
&42.9 &\textbf{39.2} &130.0 &\textbf{109.5} 
&221.9 &\textbf{153.2} &121.4 &\textbf{58.8}   \\ \cline{2-10}
&\multicolumn{1}{|@{\,}l@{\,}||}{$1$-bit} 
&947.8 &\textbf{97.4} &942.7 &\textbf{160.8} 
&866.2 &\textbf{175.8} &927.3 &\textbf{136.3} \\ \cline{2-10}
&\multicolumn{1}{|@{\,}l@{\,}||}{Robust} 
&187.3 &\textbf{42.2} &208.7 &\textbf{56.1} 
&269.1 &\textbf{79.2} &187.7 &\textbf{61.7} \\ \hline

%N = 5000
%\hline
%\multicolumn{1}{|@{\,}l@{\,}|}{AWGN} &\textbf{8.40} &290.31 &\textbf{16.24} &52.02 &\textbf{24.13} &86.93 &\textbf{15.53} &35.89   \\ \hline

%N = 10000
%\hline
%\multicolumn{1}{|@{\,}l@{\,}|}{AWGN} &\textbf{37.69} &1530.89 &\textbf{63.56} &212.39 &\textbf{107.22} &357.51 &\textbf{61.43} &141.12   \\ \hline

\end{tabular}
\egroup
}

%%%%%%%%%%%%%%%%%%%%%%%%%%%%%%%%%%%%%%%%%%%%%%%%%%%%%%%%%%%%%%%%%%%%%%%%%%%%%
\section{Conclusions} \label{sec:conc}

We proposed adaptive damping and mean-removal modifications of GAMP that help prevent divergence in the case of ``difficult'' $\vec{A}$ matrices.
We then numerically demonstrated that the resulting modifications significantly increase GAMP's robustness to non-zero-mean, low-rank product, column-correlated, and ill-conditioned $\vec{A}$ matrices.
Moreover, they provide robustness similar to the recently proposed SwAMP algorithm, whilerunning faster than the current SwAMP implementation.
For future work, we note that the sequential update of SwAMP could in principle be combined with the proposed mean-removal and/or adaptive damping to perhaps achieve a level robustness greater than either SwAMP or (M)AD-GAMP. 

%%%%%%%%%%%%%%%%%%%%%%%%%%%%%%%%%%%%%%%%%%%%%%%%%%%%%%%%%%%%%%%%%%%%%%%%%%%%%
\clearpage
\bibliographystyle{IEEEbib}
\bibliography{macros_abbrev,stc,books,blind,comm,misc,multicarrier,sparse,machine,phase}

\begin{thebibliography}{10}

\bibitem{Rangan:ISIT:11}
S.~Rangan,
\newblock ``Generalized approximate message passing for estimation with random
  linear mixing,''
\newblock in {\em Proc. IEEE Int. Symp. Inform. Thy.}, Aug. 2011, pp.
  2168--2172,
\newblock (full version at \emph{arXiv:1010.5141}).

\bibitem{Bayati:ISIT:12}
M.~Bayati, M.~Lelarge, and A.~Montanari,
\newblock ``Universality in polytope phase transitions and iterative
  algorithms,''
\newblock in {\em Proc. IEEE Int. Symp. Inform. Thy.}, Boston, MA, June 2012,
  pp. 1643--1647,
\newblock (full paper at \emph{arXiv:1207.7321}).

\bibitem{Javanmard:II:13}
Adel Javanmard and Andrea Montanari,
\newblock ``State evolution for general approximate message passing algorithms,
  with applications to spatial coupling,''
\newblock {\em Inform. Inference}, vol. 2, no. 2, pp. 115--144, 2013.

\bibitem{Rangan:ISIT:13}
S.~Rangan, P.~Schniter, E.~Riegler, A.~Fletcher, and V.~Cevher,
\newblock ``Fixed points of generalized approximate message passing with
  arbitrary matrices,''
\newblock in {\em Proc. IEEE Int. Symp. Inform. Thy.}, July 2013, pp. 664--668,
\newblock (full version at \emph{arXiv:1301.6295}).

\bibitem{Krzakala:ISIT:14}
F.~Krzakala, A.~Manoel, E.~W. Tramel, and L.~Zdeborov\'a,
\newblock ``Variational free energies for compressed sensing,''
\newblock in {\em Proc. IEEE Int. Symp. Inform. Thy.}, July 2014, pp.
  1499--1503,
\newblock (see also \emph{arXiv:1402.1384}).

\bibitem{Caltagirone:ISIT:14}
F.~Caltagirone, F.~Krzakala, and L.~Zdeborov\'a,
\newblock ``On convergence of approximate message passing,''
\newblock in {\em Proc. IEEE Int. Symp. Inform. Thy.}, July 2014, pp.
  1812--1816,
\newblock (see also \emph{arXiv:1401.6384}).

\bibitem{Rangan:ISIT:14}
S.~Rangan, P.~Schniter, and A.~Fletcher,
\newblock ``On the convergence of generalized approximate message passing with
  arbitrary matrices,''
\newblock in {\em Proc. IEEE Int. Symp. Inform. Thy.}, July 2014, pp. 236--240,
\newblock (full version at \emph{arXiv:1402.3210}).

\bibitem{Manoel:swamp:14}
Andre Manoel, Florent Krzakala, Eric~W. Tramel, and Lenka Zdeborov\'a,
\newblock ``Sparse estimation with the swept approximated message-passing
  algorithm,''
\newblock {\em arXiv:1406.4311}, June 2014.

\bibitem{GAMPmatlab}
S.~Rangan, P.~Schniter, J.~T. Parker, J.~Ziniel, J.~Vila, M.~Borgerding, and
  et~al.,
\newblock ``{GAMP}matlab,''
\newblock {\footnotesize\verb+https://sourceforge.net/projects/gampmatlab/+}.

\bibitem{Heskes:NIPS:02}
T.~Heskes,
\newblock ``Stable fixed points of loopy belief propagation are minima of the
  {B}ethe free energy,''
\newblock in {\em Proc. Neural Inform. Process. Syst. Conf.}, Vancouver, B.C.,
  Dec. 2002, pp. 343--350.

\bibitem{Schniter:ALL:12}
P.~Schniter and S.~Rangan,
\newblock ``Compressive phase retrieval via generalized approximate message
  passing,''
\newblock in {\em Proc. Allerton Conf. Commun. Control Comput.}, Monticello,
  IL, Oct. 2012, pp. 815--822,
\newblock (full version at \emph{arXiv:1405.5618}).

\bibitem{SwAMP}
Andre Manoel, Florent Krzakala, Eric~W. Tramel, and Lenka Zdeborov{\'a},
\newblock ``{SwAMP} demo user's manual,''
\newblock {\footnotesize\verb+https://github.com/eric-tramel/SwAMP-Demo+}.

\bibitem{Kamilov:TSP:12}
U.~S. Kamilov, V.~K. Goyal, and S.~Rangan,
\newblock ``Message-passing de-quantization with applications to compressed
  sensing,''
\newblock {\em IEEE Trans. Signal Process.}, vol. 60, no. 12, pp. 6270--6281,
  Dec. 2012.

\end{thebibliography}
\end{document}